\documentclass{vldb}
\usepackage{graphicx}
\usepackage{balance}  % for  \balance command ON LAST PAGE  (only there!)

\usepackage{caption}
\usepackage{amssymb} % main package.
\usepackage{amsmath} % for writing math in between text
\usepackage{mdframed} % for the algorithms frame boxes
\usepackage{comment} % allowing comments in the file

\newcommand{\intersection}{\mbox{$\cap$}}

\vldbTitle{Improved Cardinality Estimation by Learning Queries Containment Rates}
\vldbAuthors{Rojeh Hayek, Oded Shmuile}
\vldbDOI{https://doi.org/10.14778/xxxxxxx.xxxxxxx}
\vldbVolume{12}
\vldbNumber{xxx}
\vldbYear{2020}

\begin{document}
\nocite{*} % to display all the references

% ****************** TITLE ****************************************

\title{Improved Cardinality Estimation by Learning Queries Containment Rates}

% possible, but not really needed or used for PVLDB:
%\subtitle{[Extended Abstract]
%\titlenote{A full version of this paper is available as\textit{Author's Guide to Preparing ACM SIG Proceedings Using \LaTeX$2_\epsilon$\ and BibTeX} at \texttt{www.acm.org/eaddress.htm}}}

% ****************** AUTHORS **************************************

\numberofauthors{2} 

\author{
% 1st. author
\alignauthor
Rojeh Hayek\\
       \affaddr{CS Department, Technion}\\
       \affaddr{Haifa 3200003, Israel}\\
       \email{srojeh@cs.technion.ac.il}
% 2nd. author
\alignauthor
Oded Shmueli\\
       \affaddr{CS Department, Technion}\\
       \affaddr{Haifa 3200003, Israel}\\
       \email{oshmu@cs.technion.ac.il}
}

\maketitle

\begin{abstract}
The \textit{containment rate} of query $Q1$ in query $Q2$ over database $D$ is the percentage of $Q1$'s result tuples over $D$ that are also in $Q2$'s result over $D$. We directly estimate containment rates between pairs of queries over a specific database. For this, we use a specialized deep learning scheme, CRN, which is tailored to representing pairs of SQL queries. Result-cardinality estimation is a core component of query optimization. We describe a novel approach for estimating queries’ result-cardinalities using estimated containment rates among queries.  This containment rate estimation may rely on CRN or embed, unchanged, known \textit{cardinality} estimation methods. Experimentally, our novel approach for estimating cardinalities, using containment rates between queries, on a challenging real-world database, realizes significant improvements to state of the art cardinality estimation methods. 
\end{abstract}

\section{Introduction}
Query $Q1$ is contained in (resp. equivalent to), query $Q2$, analytically, if for all the database states $D$, $Q1$'s result over $D$ is contained in (resp., equals) $Q2$'s result over $D$.
Query containment is a well-known concept that has applications in query optimization. It has been extensively researched in database theory, and many algorithms were proposed for determining containment under different assumptions \cite{cnt1,cnt2,cnt3,cnt4}. However, determining query containment analytically is not practically sufficient. Two queries may be analytically unrelated by containment, although, the execution result on a \textit{specific} database of one query may actually be contained in the other. For example, consider the queries:\\
Q1: \textit{select * from movies where title = 'Titanic'}\\
Q2: \textit{select * from movies where release = 1997 and director = 'James Cameron'}\\
Both queries execution results are identical since there is only one movie called Titanic that was released in 1997 and directed by James Cameron (he has not directed any other movie in 1997). Yet, using the analytic criterion, the queries are unrelated at all by containment.

To our knowledge, while query containment and equivalence have been well researched in past decades, determining the containment rate between two queries on a \textit{specific} database, has not been considered by past research. 

By definition, the containment rate of query $Q1$ in query $Q2$ on database $D$ is the percentage of rows in $Q1$'s execution result over $D$ that are also in $Q2$'s execution result over $D$. Determining containment rates allows us to solve other problems, such as determining equivalence between two queries, or whether one query is fully contained in another, on the same \textit{specific} database. In addition, containment rates can be used in many practical applications, for instance, query clustering, query recommendation \cite{SimTuples,SimStructure}, and in cardinality estimation as will be described subsequently.

Our approach for estimating containment rates is based on a specialized deep learning model, CRN, which enables us to express query features using sets and vectors. An input query is converted into three sets, $T$, $J$ and $P$ representing the query's tables, joins and column predicates, respectively. Each element of these sets is represented by a vector. Using these vectors, CRN generates a single vector that represents the whole input query. Finally, to estimate the containment rate of two represented queries, CRN measures the distance between the representative vectors of both queries, using another specialized neural network. Thus, the CRN model relies on the ability of the neural network to learn the vector representation of queries relative to the \textit{specific} database. As a result, we obtain a small and accurate model for estimating containment rates.

In addition to the CRN model, we introduce a novel technique for estimating queries' cardinalities using estimated query containment rates. We show that using the proposed technique we improve current cardinality estimation techniques significantly. This is especially the case when there are multiple joins, where the known cardinality estimation techniques suffer from under-estimated results and errors that grow exponentially as the number of joins increases \cite{joinsHard}. Our technique estimates the cardinalities more robustly (x150/x175 with 4 joins queries, and x1650/x120 with 5 joins queries, compared with PostgreSQL and MSCN, respectively). We compare our technique with PostgreSQL \cite{postgreSQL}, and the pioneering MSCN model \cite{LearnedCrd}, by examining, on the real-world IMDb database \cite{HowGoodCar}, join crossing correlations queries which are known to present a tough challenge to cardinality estimation methods \cite{HowGoodCar,crdHard,JoinCross}.

We show that by employing known existing cardinality estimation methods for containment estimation, we can improve on their cardinality estimates as well, without changing the methods themselves. Thus, our novel approach is highly promising for solving the cardinality estimation problem, the "Achilles heel" of query optimization \cite{crdHard2}, a cause of many performance issues \cite{HowGoodCar}.

The rest of this paper is organized as follows. In Section \ref{Containment Rate Definition} we define the containment rate problem and in Sections \ref{Learned Containment Rates}-\ref{Containment Evaluation} we describe and evaluate the CRN model for solving this problem. In Sections \ref{Cardinality Estimation Using Containment Rates}-\ref{Cardinality Evaluation} we describe and evaluate our new approach for estimating cardinalities using containment rates. In Section \ref{Improving Existing Cardinality Estimation Models} we show how one can adapt the new ideas to improve existing cardinality estimation models. Sections \ref{Related work}-\ref{Conclusion} present related work, conclusions and future work. 

\section{Containment Rate Definition}
\label{Containment Rate Definition}
We define the containment rate between two queries $Q1$, and $Q2$ on a \textit{specific} database $D$. \textit{Query $Q1$ is $x\%$-contained in query $Q2$ on database $D$ if precisely $x\%$ of $Q1$'s execution result rows on database $D$ are also in $Q2$'s execution result on database $D$.} The containment rate is formally a function from \textbf{QxQxD} to \textbf{R}, where \textbf{Q} is the set of all queries, \textbf{D} of all databases, and \textbf{R} the Real numbers. This function can be directly calculated using the cardinality of the results of queries $Q1$ and $Q2$ as follows:
$$ x\% = \frac{|Q1(D)\ \intersection\ Q2(D)|}{|Q1(D)|} * 100 $$
Where, $Q(D)$ denotes $Q$'s execution result on database $D$. (in case $Q1$'s execution result is empty, then $Q1$ is 0\%-contained in $Q2$). Note that the containment rate is defined only on pairs of queries whose SELECT and FROM clauses are \textit{identical}. 

\subsection{Containment Rate Operator}
We denote the containment rate \textit{operator} between queries $Q1$ and $Q2$ on database $D$ as:
$$ Q1 \subset_{\%}^D Q2 $$
Operator $\subset_{\%}^D$ returns the containment rate between the given input queries on database $D$. That is, $Q1 \subset_{\%}^D Q2$ returns $x\%$, if $Q1$ is $x\%$-contained in query $Q2$ on database $D$. For simplicity, we do not mention the \textit{specific} database, as it is usually clear from context. Therefore, we write the containment rate operator as $\subset_{\%}$.

\section{Learned Containment Rates}
\label{Learned Containment Rates}
From a high-level perspective, applying machine learning to the containment rate estimation problem is straightforward. Following the training of the CRN model with pairs of queries $(Q1,Q2)$ and the actual containment rates $Q1 \subset_{\%} Q2$, the model is used as an estimator for other, unseen pairs of queries. There are, however, several questions whose answers determine whether the machine learning model (CRN) will be successful. (1) Which supervised learning algorithm/model should be used. (2) How to represent queries as input and the containment rates as output to the model ("featurization"). (3) How to obtain the initial training dataset ("cold start problem"). Next, we describe how we address each one of these questions.

\subsection{Cold Start Problem}
\label{Defining the database and the development set}
\subsubsection{Defining the Database}
\label{Defining the database}
We generated a training-set, and later on evaluated our model on it, using the IMDb database. IMDb contains many correlations and has been shown to be very challenging for cardinality estimators \cite{HowGoodCar}. This database contains a plethora of information about movies and related facts about actors, directors, and production companies, with more than 2.5M movie titles produced over 130 years (starting from 1880) by 235,000 different companies with over 4M actors.

\subsubsection{Generating the Development Dataset}
\label{queries generator}
Our approach for solving the "cold start problem" is to obtain an initial training corpus using a specialized queries generator that randomly generates queries based on the IMDB schema and the actual columns values. Our queries generator generates the dataset in three main steps. In the first step, it repeatedly generates multiple SQL queries as follows. It randomly chooses a set of tables $t$, that can join with each other in the database. Then, it adds the corresponding join edges to the query. For each base table $bt$ in the chosen set of tables $t$, it uniformly draws the number of query predicates $p_{bt}$ (0 $\leq p_{bt} \leq$ number of non-key columns in table $bt$). Subsequently, for each predicate it uniformly draws a non-key column from the relevant table $bt$, a predicate type $(<,\ =,\ or\ >)$, and a value from the corresponding column values range in the database.
To avoid a combinatorial explosion, and to simplify the problem that the model needs to learn, we force the queries generator to create queries with up to two joins and let the model generalize to a larger number of joins. Note that all the generated queries include a SELECT * clause. They are denoted as \textit{initial-queries}. 

To create pairs of queries that are contained in each other with different containment rates, we generate, in the second step, queries that are "similar" to the \textit{initial-queries}, but still, different from them, as follows. For each query $Q$ in \textit{initial-queries}, the generator repeatedly creates multiple queries by randomly changing query $Q$'s predicates' types, or the predicates' values, and by randomly adding additional predicates to the original query $Q$. This way, we create a "hard" dataset, which includes pairs of queries that look "similar", but having mutual containment rates that vary significantly. Finally, in the third and last step, using the queries obtained from both previous steps, the queries generator generates pairs of queries whose FROM clauses are identical. (Note that our queries generator's first step is similar to MSCN's generator \cite{LearnedCrd}, however, in the second step we create more complicated queries).

After generating the dataset, we execute the dataset queries on the IMDb database, to obtain their true containment rates. Using this process, we obtain an initial training set for our model, which consists of 100,000 pairs of queries with zero to two joins. We split the training samples into 80\% training samples and 20\% validation samples.

\subsection{Model}
\label{Model}
Featurizing all the queries' literals and predicates as one "big hot vector", over all the possible words that may appear in the queries, is impractical. Also, serializing the queries' SELECT, FROM, and WHERE clauses elements into an ordered sequence of elements, is not practical, since the order in these clauses is arbitrary. 

Thus, standard deep neural network architectures such as simple multi-layer perceptrons \cite{RNNMLPCNN}, convolutional neural networks \cite{RNNMLPCNN}, or recurrent neural networks \cite{RNNMLPCNN}, are not directly proper to our problem.

\begin{figure}[h!]
\begin{center}
  \includegraphics[width=\linewidth]{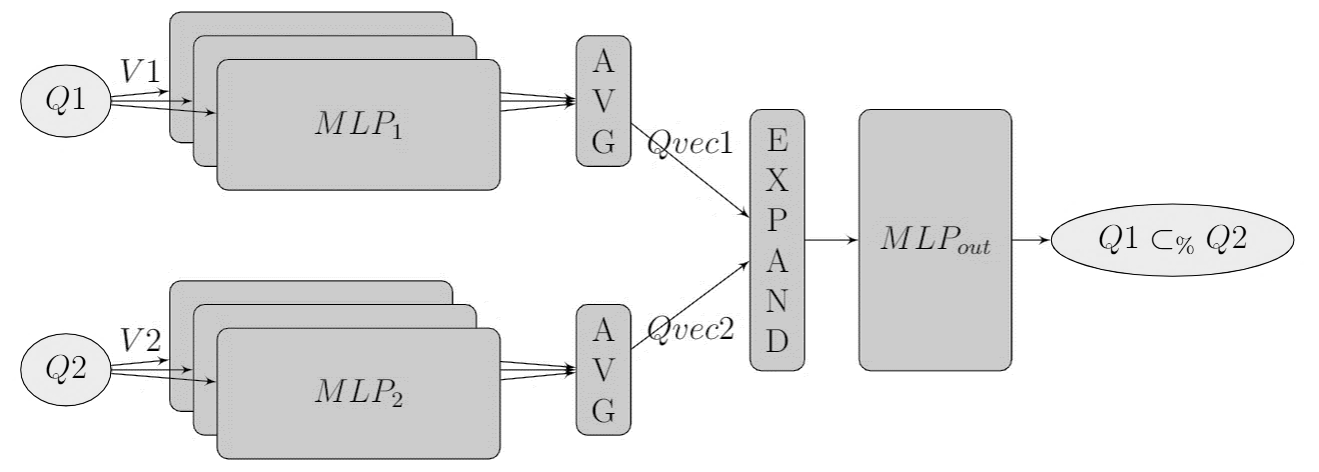}
  \caption{CRN Model Archeticture.}
  \label{CRN Model Archeticture}
\end{center}
\end{figure}

Our \textit{Containment Rate Network} (CRN) model uses a specialized vector representation for representing the input queries and the output containment rates. As depicted in Figure \ref{CRN Model Archeticture}, the CRN model runs in three main stages. Consider an input queries pair $(Q1,Q2)$. In the first stage, we convert $Q1$ (resp., $Q2$) into a set of vectors $V1$ (resp., $V2$). Thus $(Q1,Q2)$ is represented by $(V1,V2)$. In the second stage, we convert set $V1$ (resp., $V2$) into a unique single representative vector $Qvec1$ (resp., $Qvec2$), using a specialized neural network, $MLP_i$, for each set separately. In the third stage, we estimate the containment rate $Q1 \subset_{\%} Q2$, using the representative vectors $Qvec1$ and $Qvec2$, and another specialized neural network, $MLP_{out}$.

\begin{figure*}[h!]
\begin{center}
  \includegraphics[width=\linewidth]{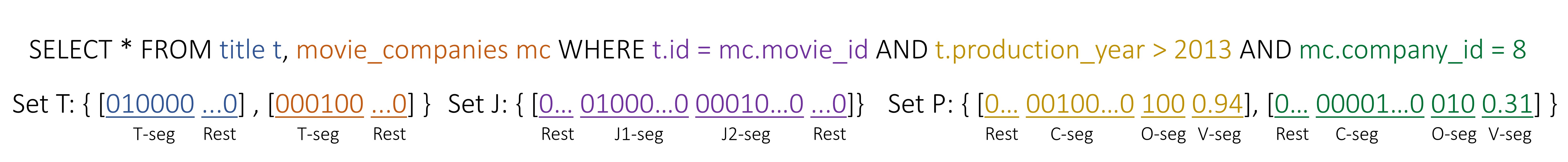}
  \caption{Query featurization as sets of feature vectors obtained from sets $T$, $J$ and $P$ (Rest denotes zeroed portions of a vector).}
  \label{example}
\end{center}
\end{figure*} 

\subsubsection{First Stage, from $(Q1,Q2)$ to $(V1,V2)$}
In the same way as MSCN model \cite{LearnedCrd}, we represent each query $Q$ as a collection of three sets $(T,J,P)$. $T$ is the set of all the tables in $Q$'s FROM clause. $J$ is the set of all the joins (i.e., join clauses) in $Q$'s WHERE clause. $P$ is the set of all the (column) predicates in $Q$'s WHERE clause. Using sets $T$, $J$, and $P$, we obtain a set of vectors $V$ representing the query, as described later.
Unlike MSCN, in our model all the vectors of set $V$ have the same dimension and the same segmentation as depicted in Table \ref{table:Vector Segmentation}, where 
$\#T$ is the number of all the tables in the database, $\#C$ is the number of all the columns in all the database tables, and $\#O$ is the number of possible predicates operators. In total, the vector dimension is $\#T + 3*\#C + \#O + 1$, denoted as $L$. 

The queries tables, joins and column predicates (sets $T$, $J$ and $P$) are inseparable, hence, treating each set individually using different neural networks may disorientate the model. Therefore, we choose to featurize these sets using the same vector format in order to ease learning. 

\begin{table}[h!]
\centering
\resizebox{\linewidth}{!}{
\begin{tabular}{c|c|c|c|c|c|c}
\hline
\hline
$\bold{Type}$ & $\bold{Table}$ & \multicolumn{2}{c|}{$\bold{Join}$} & \multicolumn{3}{c}{$\bold{Column\ Predicate}$} \\
\hline
$\bold {Segment}$ & \textbf{T-seg} & \textbf{J1-seg} & \textbf{J2-seg} &  \textbf{C-seg} & \textbf{O-seg} & \textbf{V-seg} \\
\hline
$\bold {Segment\ size}$ & $\#T$ & $\#C$ &  $\#C$ &  $\#C$ & $\#O$ & $1$ \\
\hline
$\bold {Featurization}$ & $one\ hot$ & $one\ hot$ & $one\ hot$ &  $one\ hot$ & $one\ hot$ & $norm.$ \\
\hline
\hline
\end{tabular}
}
\caption{Vector Segmentation.}
\label{table:Vector Segmentation}
\end{table}

Element of sets $T$, $J$, and $P$, are represented by vectors as follows (see a simple example in Figure \ref{example}). All the vectors have the same dimension $L$. 
Each table $t \in T$ is represented by a unique one-hot vector (a binary vector of length $\#T$ with a single non-zero entry, uniquely identifying a specific table) placed in the T-seg segment. 
Each join clause of the form $(col1,=,col2) \in J$ is represented as follows. $col1$ and $col2$ are represented by a unique one-hot vectors placed in J1-seg and J2-seg segments, respectively.
Each predicate of the form $(col,op,val) \in P$ is represented as follows. $col$ and $op$ are represented by a unique one-hot vectors placed in the C-seg and V-seg segments, respectively. $val$ is represented as a normalized value $\in [0, 1]$, normalized using the minimum and maximum values of the respective column, placed in the V-seg segment. For each vector, all the other unmentioned segments are zeroed. Given input queries pair, $(Q1,Q2)$, we convert query $Q1$ (resp., $Q2$) into sets $T$, $J$ and $P$, and \textit{each} element of these sets is represented by a vector as described above, together generating set $V1$ (resp., $V2$).

\subsubsection{Second Stage, from $(V1,V2)$ to $(Qvec1,Qvec2)$}
Given set of vectors $V_i$, we present each vector of the set into a fully-connected one-layer neural network, denoted as $MLP_i$, converting each vector into a vector of dimension $H$. The final representation $Qvec_i$ for this set is then given by the average over the individual transformed representations of its elements, i.e.,
$$ Qvec_i = \frac{1}{|V_i|} \sum_{v \in V_i} MLP_i (v)$$
$$ MLP_i(v) = Relu(vU_i + b_i)$$
Where $U_i \in R^{LxH}$, $b_i \in R^{H}$ are the learned weights and bias, and $v \in R^{L}$ is the input row vector. We choose an average (instead of, e.g., sum) to ease generalization to different numbers of elements in the sets, as otherwise the overall magnitude of $Qvec$ would vary depending on the number of elements in the set $V_i$.

\subsubsection{Third Stage, from $(Qvec1,Qvec2)$ to $Q1 \subset_{\%} Q2$}
Given the representative vectors of the input queries,\\$(Qvec1,Qvec2)$, we aim to predict the containment rate $Q1 \subset_{\%} Q2$ as accurately as possible. Since we do not know what a "natural" distance measure is in the representative queries vector space, encoded by the neural networks of the second step, we use a fully-connected two-layer neural network, denoted as $MLP_{out}$, to compute the estimated containment rate of the input queries, leaving it up to this neural network to learn the correct distance function. $MLP_{out}$ takes as input a vector of size $4H$ which is constructed from $Qvec1$ and $Qvec2$. The first layer converts the input vector into a vector of size $2H$. The second layer converts the obtained  vector of size $2H$, into a single value representing the containment rate.
$$ \hat{y} = MLP_{out}(Expand(Qvec1,Qvec2)) $$
$$ MLP_{out}(v) = Sigmoid(ReLU(vU_{out1} + b_{out1})U_{out2} + b_{out2}) $$
$$ Expand(v_1,v_2) = [v_1,\ \ v_2,\ \ abs(v_1 - v_2),\ \ v_1 \odot v_2]$$
Here, $\hat{y}$ is the estimated containment rate, $U_{out1} \in R^{4Hx2H}$,  $b_{out1} \in R^{2H}$ and $U_{out2} \in R^{2Hx1}$,  $b_{out2} \in R^{1}$ are the learned weights and bias, and $\odot$ is the dot-product function. 

In order to estimate the containment rates more accurately, we use the $Expand$ function which creates a row concatenated vector of size $4H$ using vectors $Qvec1$ and $Qvec2$.

We use the $ReLU$\footnote{ReLU(x) = max(0,x); see \cite{ActivationFunc}.} activation function for hidden layers in all the neural networks, as they show strong empirical performance advantages and are fast to evaluate.

In the final step, we apply the $Sigmoid$\footnote{Sigmoid(x) = $1/(1+e^{-x})$; see \cite{ActivationFunc}.} activation function in the second layer to output a float value in the range [0,1], as the containment rate values are within this interval. Therefore, we do not apply any featurization on the containment rates (the output of the model) and the model is trained with the actual containment rate values without any featurization steps.

\subsubsection{Loss Function} 
Since we are interested in minimizing the ratio between the predicted and the actual containment rates, we use the q-error metric in our evaluation. We train our model to minimize the mean q-error \cite{qerror}, which is the ratio between an estimated and the actual contaminate rate (or vice versa). Let $y$ be the true containment rate, and $\hat{y}$ the estimated rate, then the q-error is defined as follows.
$$ q-error(y,\hat{y})\ =\ \hat{y} > y\ ?\ \frac{\hat{y}}{y}\ :\ \frac{y}{\hat{y}}$$

In addition to optimizing the mean q-error, we also examined the mean squared error (MSE) and the mean absolute error (MAE) as optimization goals. MSE and MAE would optimize the squared/absolute differences between the predicted and the actual containment rates. Optimizing with theses metrics makes the model put less emphasis on heavy outliers (that lead to large errors). Therefore, we decided to optimize our model using the q-error metric which yielded better results.

\subsection{Training and Testing Interface}
 Building CRN involves two main steps. (1) Generating a random training set using the schema and data information as described in Section \ref{Defining the database and the development set}. (2) Repeatedly using this training data, we train the CRN model as described in Section \ref{Model} until the mean q-error of the validation test starts to converges to its best absolute value. That is, we use the early stopping technique \cite{earlyStopping} and stop the training before convergence to avoid over-fitting. Both steps are performed on an immutable snapshot of the database.

After the training phase, to predict the containment rate of an input query pair, the queries first need to be transformed into their feature representation, and then they are presented as input to the model, and the model outputs the estimated containment rate (Section \ref{Model}).

We train and test our model using the Tensor-Flow framework \cite{tensorFlow}, and make use of the efficient Adam optimizer \cite{adam} for training the model.

\subsection{Hyperparameter Search}
\label{Hyperparameter Search}
To optimize our model's performance, we conducted a search over its hyperparameter space. In particular, we focused on tuning the neural networks hidden layer size (H). 

\begin{figure}[h!]
\begin{center}
  \includegraphics[width=\linewidth]{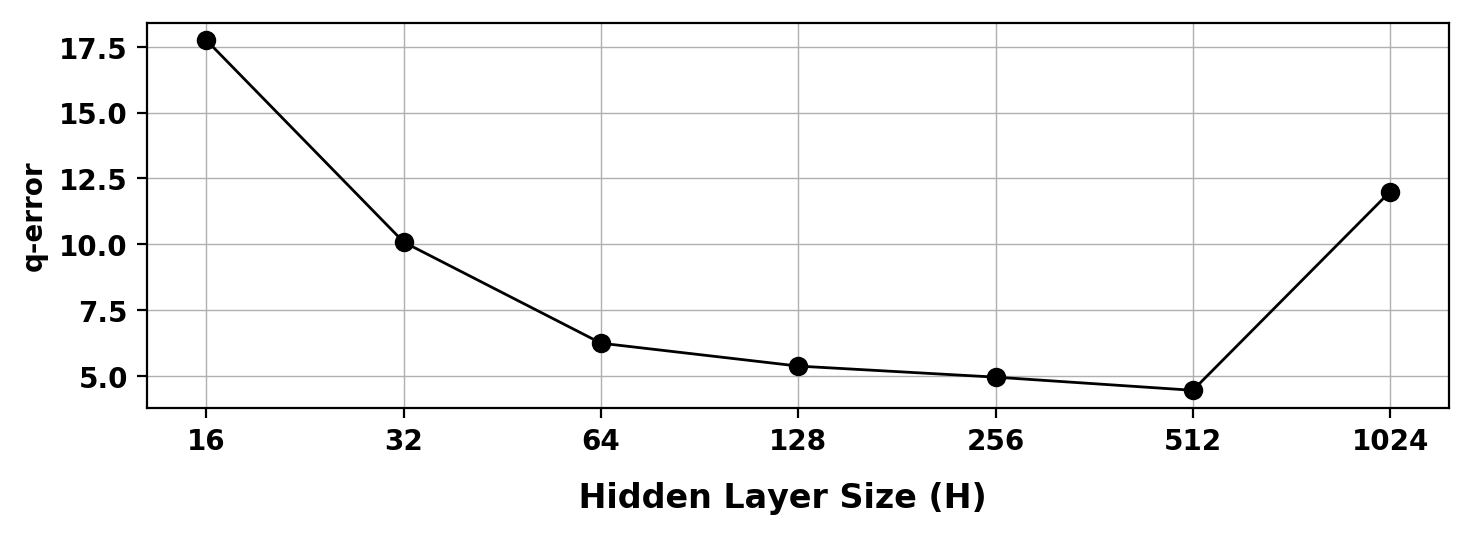}
  \caption{The mean q-error on the validation set with different hidden layer sizes.}
  \label{The mean q-error on the validation set with different hidden layer sizes}
\end{center}
\end{figure}

Note that the same H value is shared in all the neural networks of the CRN model, as described in section \ref{Model}. 
During the tuning of the size hyperparameter of the neural network hidden layer, we found that increasing the size of our hidden layer generally led to an increase in the model accuracy, till it reached the best mean q-error on the validation test. Afterwards, the results began to decline in quality because of over-fitting (See Figure \ref{The mean q-error on the validation set with different hidden layer sizes}). Hence, we choose a hidden layer of size 512, as a good balance between accuracy and training time. Overall, we found that our model performs similarly well across a wide range of settings when considering different batch sizes and learning rates.

\subsection{Model Computational Costs}
We analyze the training, prediction, and space costs of the CRN model with the default hyperparameters (H=512, batch size=128, learning rate=0.001). 

\subsubsection{Training Time}
\begin{figure}[h!]
\begin{center}
  \includegraphics[width=\linewidth]{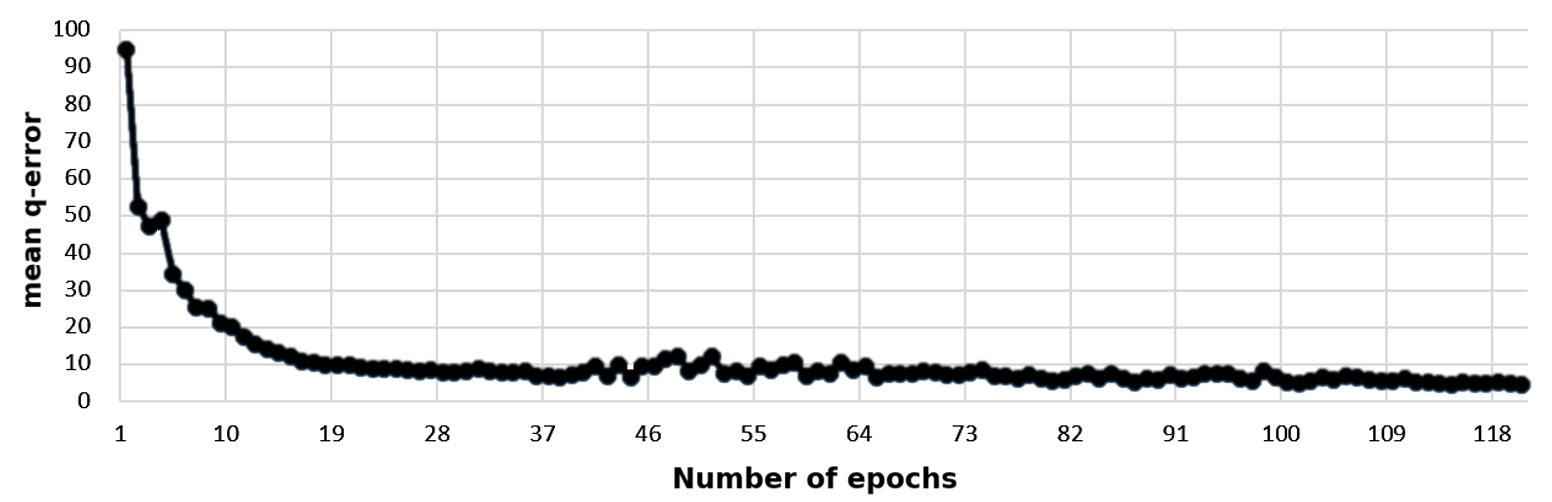}
  \caption{Convergence of the mean q-error on the validation set.}
  \label{Convergence of the mean q-error on the validation set}
\end{center}
\end{figure}

Figure \ref{Convergence of the mean q-error on the validation set} shows how the mean q-error of the validation set decreases with additional epochs, until convergence to a mean q-error of around 4.5. The CRN model requires almost 120 passes on the training set to converge. On average, measured across six runs, a training run with 120 epochs takes almost 200 minutes.

\subsubsection{Prediction Time}
The prediction process is dominated by converting the input queries into the corresponding vectors, and then presenting these vectors as input to the CRN model. On average, the prediction time is 0.5ms per single pair of queries, including the overhead introduced by the Tensor-Flow framework.

\subsubsection{Model Size}
The CRN model includes all the learned parameters mentioned in Section \ref{Model} ($U_1$, $U_2$, $U_{out1}$, $U_{out2}$, $b_1$, $b_2$, $b_{out1}$, $b_{out2}$). In total, there are $2*L*H + 8*H^2 + 6*H +1$ learned parameters. In practice, the size of the model, when serialized to disk, is roughly $1.5MB$.

\section{Containment Evaluation}
\label{Containment Evaluation}
In this section we describe how we compared the CRN model to other (baseline) methods. Since to the best of our knowledge, the problem of determining containment rate has not been addressed till now, we used a transformation as described in Section \ref{From Cardinality to Containment} below.

\subsection{From Cardinality to Containment}
\label{From Cardinality to Containment}
To our knowledge, this is the first work to address the problem of containment rate estimation. In order to compare our results with different baseline methods, we used existing cardinality estimation methods to predict the containment rates, using the Crd2Cnt transformation, as depicted in the middle part diagram in Figure \ref{fig:teaser}.

\subsubsection{The Crd2Cnt Transformation}
\label{Crd2Cnt Transformation}
Given a cardinality estimation model\footnote{Here "model" may refer to an ML model or simply to a method.} $M$, we can convert it to a containment rate estimation model using the Crd2Cnt transformation which returns a model $M'$ for estimating containment rates. The obtained model $M'$ functions as follows. Given input queries $Q1$ and $Q2$, whose containment rate $Q1 \subset_{\%} Q2$ needs to be estimated:
\begin{itemize}
    \item Calculate the cardinality of query $Q1\intersection Q2$ using $M$.
    \item Calculate the cardinality of query $Q1$ using $M$.
    \item Then, the containment rate estimate is: $$Q1 \subset_{\%} Q2 = \frac{|Q1\intersection Q2|}{|Q1|}$$
\end{itemize}

Here, $Q1\intersection Q2$ is the intersection query of $Q1$ and $Q2$ whose SELECT and FROM clauses are identical to $Q1$'s (or $Q2$'s) clauses, and whose WHERE clause is $Q1$'s AND $Q2$'s WHERE clauses. Note that, by definition, if $|Q1|=0$ then $Q1 \subset_{\%} Q2=0$.

Given model $M$, we denote the obtained model $M'$, via the Crd2Cnt transformation, as Crd2Cnt($M$).

We compared the CRN model predictions to those based on the other examined cardinality estimation models, using the Crd2Cnt transformation. We examined the PostgreSQL version 11 cardinality estimation component \cite{postgreSQL}, a simple and commonly used method for cardinality estimation. In addition, we examined the MSCN model \cite{LearnedCrd}. MSCN was shown to be superior to the best methods for estimating cardinalities such as Random Sampling (RS) \cite{RS1,RS2} and the state-of-the-art Index-Based Join Sampling (IBJS) \cite{IBJS}.

\subsubsection{Comparing with MSCN}
\label{Comparing with MSCN}
In order to make a fair comparison between the CRN model and the MSCN model, we train the MSCN model with the \textit{same} data that was used to train the CRN model.

The CRN model takes two queries as input, whereas the MSCN model takes one query as input. Therefore, to address this issue, we created the training dataset for the MSCN model as follows. For each pair of queries $(Q1,Q2)$ used in training the CRN model, we added the following two input queries to the MSCN training set:
\begin{itemize}
    \item $Q1 \intersection Q2$, along with its actual cardinality. 
    \item $Q1$, along with its actual cardinality.
\end{itemize}

Finally, we ensure that the training set includes only unique queries without repetition. This way, we both models, MSCN and CRN, are trained with the \textit{same} information.

\subsubsection{Comparing with PostgreSQL}
Comparing with PostgreSQL does not require generating an appropriate training set, since the PostgreSQL cardinality estimation component is based on database profiling techniques and does not require training.

\subsection{Evaluation Workloads}
We evaluate CRN on the IMDb dataset as described in Section \ref{Defining the database}, using two different query workloads:
\begin{itemize}
    \item cnt\_test1, a synthetic workload generated by the same queries generator as the one used for generating the training data (using a different random seed) with 1200 unique query pairs, with zero to two joins.
    \item cnt\_test2, a synthetic workload generated by the same queries generator as the one used for generating the training data (using a different random seed) with 1200 unique query pairs, with zero to \textit{five} joins. This dataset is used to examine how CRN generalizes to additional joins.
\end{itemize}

\begin{table}[h!]
\centering
\resizebox{\linewidth}{!}{
\begin{tabular}{lccccccc}
\hline
\hline
$\bold {number\ of\ joins}$ & $\bold {0}$ & $\bold {1}$ & $\bold {3}$ & $\bold {3}$ & $\bold {4}$ & $\bold {5}$ & $\bold {overall}$ \\
\hline
$\bold {cnt\_test1}$ & $400$ & $400$ & $400$ & $0$ & $0$ & $0$ & $1200$\\
$\bold {cnt\_test2}$ & $200$ & $200$ & $200$ & $200$ & $200$ & $200$ & $1200$\\
\hline
\hline
\end{tabular}
}
\caption{Distribution of joins.}
\end{table}

\subsection{The Quality of Estimates}
\label{mylabel}
\begin{figure}[h!]
\begin{center}
  \includegraphics[width=\linewidth]{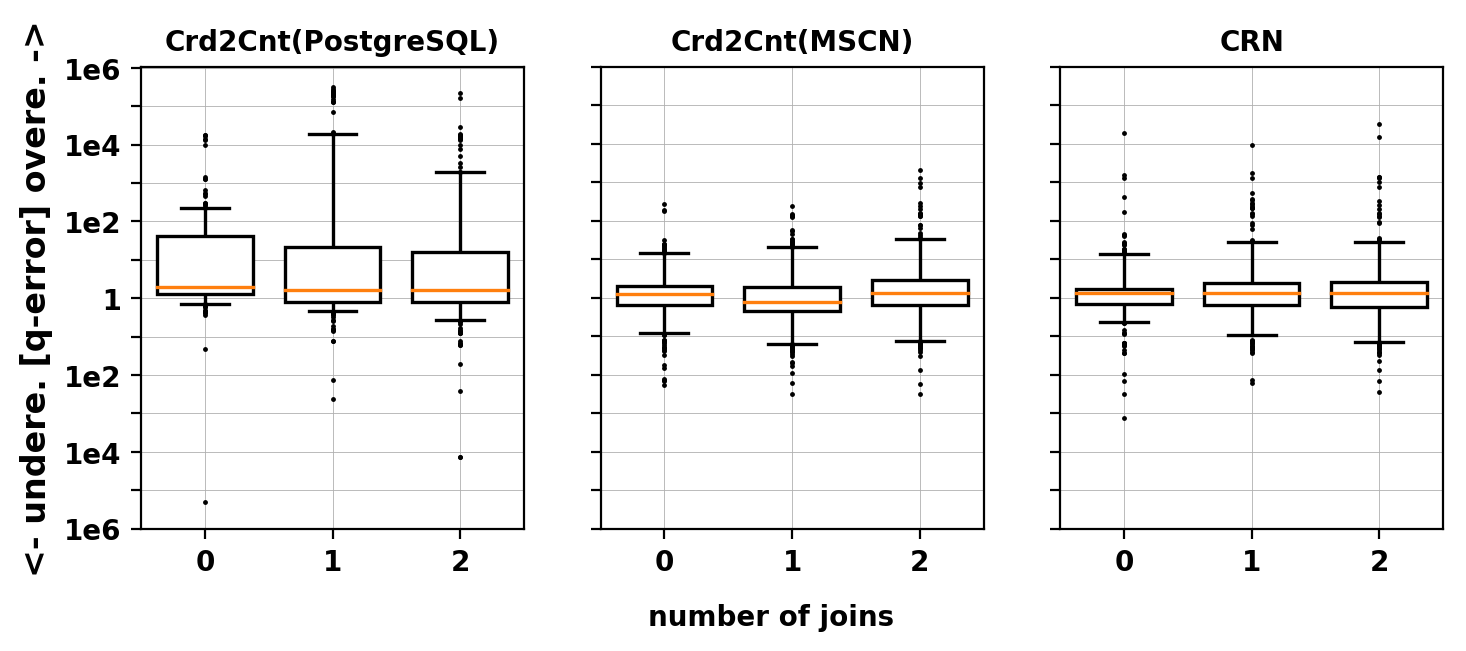}
  \caption{Estimation errors on the cnt\_test1 workload. In all the similar plots presented in this paper, the box boundaries are at the 25th/75th percentiles and the horizontal lines mark the 5th/95th percentiles. Hence, 50\% of the tests results are located within the box boundaries, and 90\% are located between the horizontal lines. The orange horizontal line mark the 50th percentile. }
  \label{fig:cnt_test1}
\end{center}
\end{figure}

Figure \ref{fig:cnt_test1} depicts the q-error of the CRN model compared to the Crd2Cnt(PostgreSQL) and Crd2Cnt(MSCN) models on the cnt\_test1 workload. While Crd2Cnt(PostgreSQL)'s errors are more skewed towards the positive spectrum,\\ Crd2Cnt(MSCN) performs extremely well as does the CRN model. Observe that we make sure to train MSCN in such a way that it will predict containment rates efficiently, while the primary purpose of the MSCN model, as described in \cite{LearnedCrd} is estimating cardinalities. That is, had we trained the MSCN model for its main purpose, with "independent" queries, we might have ended up with worse results for MSCN.

To provide a fuller picture, we also show the percentiles, maximum, and mean q-errors. As depicted in Table \ref{table:cnt_test1}, CRN provides the best results in 75\% of the tests, whereas MSCN is more robust in the margins, resulting in a better mean.

\begin{table}[h!]
\centering
\resizebox{\linewidth}{!}{
\begin{tabular}{lccccccc}
\hline
\hline
$\bold {}$ & $\bold {50th}$ & $\bold {75th}$ & $\bold {90th}$ & $\bold {95th}$ & $\bold {99th}$ & $\bold {max}$ & $\bold {mean}$\\
\hline
$\bold {Crd2Cnt(PostgreSQL)}$ & $3.5$ & $41.18$ & $365$ & $3399$ & $268745$ & $493474$ & $5492$ \\
$\bold{Crd2Cnt(MSCN)}$ & $2.84$ & $7.38$ & $\bold{19.95}$ & $\bold{41.43}$ & $\bold{274}$ & $\bold{3258}$ & $\bold{17.08}$\\
$\bold {CRN}$ & $\bold{2.52}$ & $\bold{6.17}$ & $23.04$ & $44.85$ & $991$ & $51873$ & $111$\\
\hline
\hline
\end{tabular}
}
\caption{Estimation errors on the cnt\_test1 workload. In all the similar tables presented in this paper, we provide the percentiles, maximum, and the mean q-errors of the tests. The \textit{p'}th percentile, is the q-error value below which \textit{p}\% of the test q-errors are found.}
\label{table:cnt_test1}
\end{table}

\subsection{Generalizing to Additional Joins}
\label{Generalizing to More Joins}
\begin{figure}[h!]
\begin{center}
  \includegraphics[width=\linewidth]{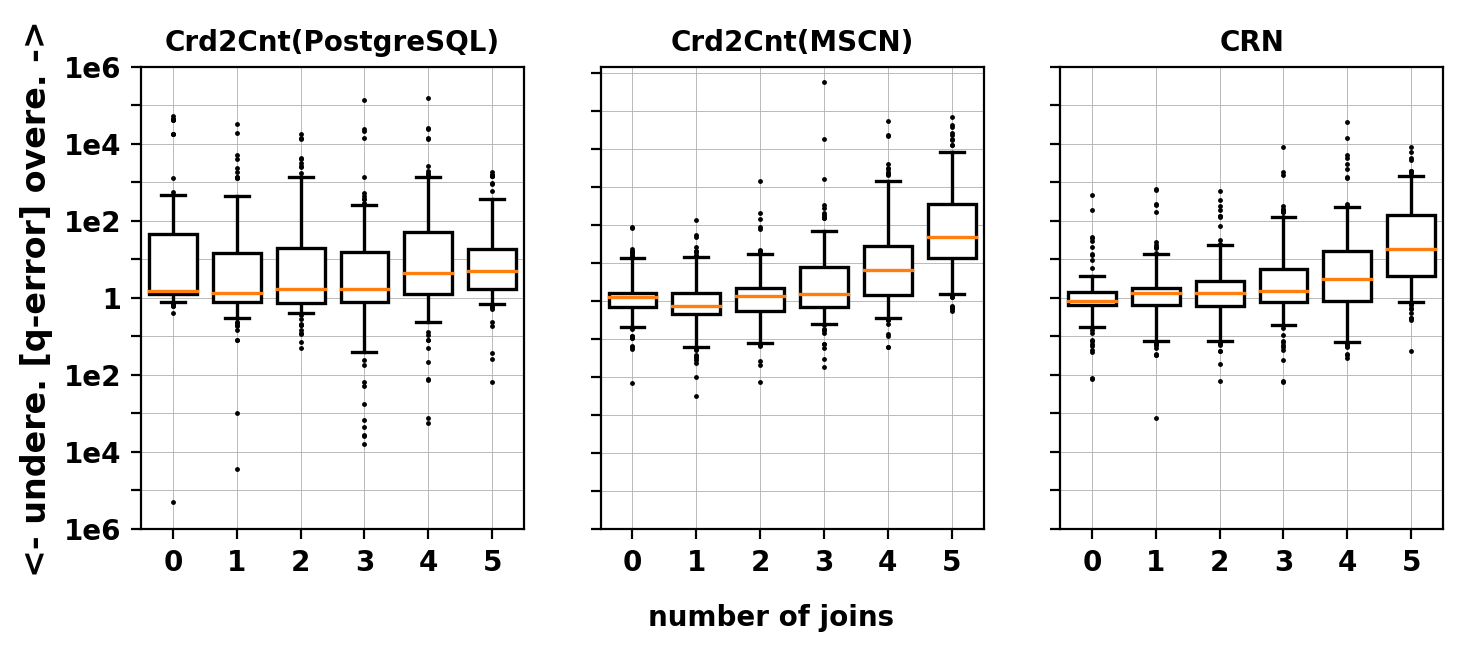}
  \caption{Estimation errors on the cnt\_test2 workload.}
  \label{fig:cnt_test2}
\end{center}
\end{figure}

In this section we examine how the CRN model generalizes to queries with a higher number of joins without having seen such queries during training. To do so, we use the crd\_test2 workload which includes queries with zero to \textit{five} joins. Recall that we trained both the CRN model and the MSCN model only with query pairs that have between zero and two joins. Examining the results, described in Table \ref{table:cnt_test2} and Figure \ref{fig:cnt_test2}, the CRN model is noticeably more robust in generalizing to queries with additional joins. The mean q-error of the CRN model is smaller by a factor of almost 8 than the mean q-errors of the other models.

\begin{table}[h!]
\centering
\resizebox{\linewidth}{!}{
\begin{tabular}{lccccccc}
\hline
\hline
$\bold {}$ & $\bold {50th}$ & $\bold {75th}$ & $\bold {90th}$ & $\bold {95th}$ & $\bold {99th}$ & $\bold {max}$ & $\bold {mean}$ \\
\hline
$\bold {Crd2Cnt(PostgreSQL)}$ & $4.5$ & $46.22$ & $322$ & $1330$ & $39051$ & $316122$ & $1345$\\
$\bold {Crd2Cnt(MSCN)}$ & $4.1$ & $17.85$ & $157$ & $754$ & $14197$ & $768051$ & $1238$\\
$\bold{CRN}$ & $\bold{3.64}$ & $\bold{13.19}$ & $\bold{96.6}$ & $\bold{255}$ & $\bold{2779}$ & $\bold{56965}$ & $\bold{161}$ \\
\hline
\hline
\end{tabular}
}
\caption{Estimation errors on the cnt\_test2 workload.}
\label{table:cnt_test2}
\end{table}

\begin{figure*}
  \includegraphics[width=\linewidth]{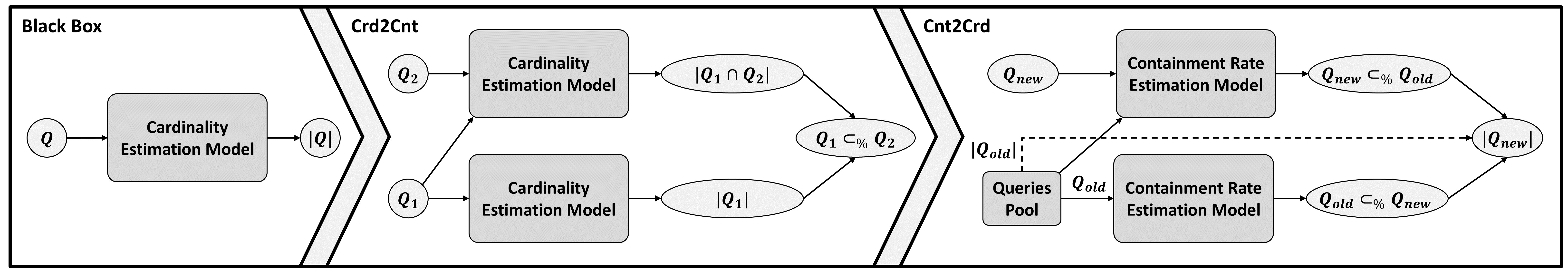}
  \caption{A novel approach, from cardinality estimation to containment rate estimation, and back to cardinality estimation by using a queries pool.}
  \label{fig:teaser}
\end{figure*}

\section{Cardinality Estimation Using\\ Containment Rates}
\label{Cardinality Estimation Using Containment Rates}

In this section we consider one application of the proposed containment rate estimation model: cardinality estimation. We introduce a novel approach for estimating cardinalities using query containment rates, and we show that using the proposed approach, we improve cardinality estimations significantly, especially in the case when there are multiple joins.

Since a traditional query optimizer is crucially dependent on cardinality estimation, which enables choosing among different plan alternatives by using the cardinality estimation of intermediate results within query execution plans. Therefore, the query optimizer must use reasonably good estimates. However, estimates produced by all widely-used database cardinality estimation models are routinely significantly wrong (under/over-estimated), resulting in not choosing the best plans, leading to slow executions \cite{HowGoodCar}.

Two principal approaches for estimating cardinalities have emerged. (1) Using database profiling \cite{postgreSQL}. (2) Using sampling techniques \cite{RS1,RS2,IBJS}. Recently, deep learning neural networks were also used for solving this problem \cite{LearnedCrd,LearnedCrd2}. However, all these approaches, with all the many attempts to improve them, have conceptually addressed the problem \textit{directly} in the same way, as a black box, where the input is a query, and the output is its cardinality estimation, as described in the leftmost diagram in Figure \ref{fig:teaser}. In our proposed approach, we address the problem differently, and we obtain better estimates as described in Section \ref{Cardinality Evaluation}.

In today's databases, the answer to a previous query is rarely useful for speeding up new queries, and the work performed in answering past queries is often ignored afterwards. Using the CRN model for predicting containment rates, we are able to change this by revealing the underlying relations between the new queries and the previous ones. 

Our new technique for estimating cardinalities mainly relies on two key ideas. The first one is the new framework in which we solve the problem. The second is the use of a \textit{queries pool} that maintains multiple previously executed queries along with their actual cardinalities, as part of the database meta information. The queries pool provides new information that enables our technique to achieve better estimates. Using a containment rate estimation model, we make use of previously executed queries along with their actual cardinalities to estimate the result-cardinality of a new query. This is done with the help of a simple transformation from the problem of containment rate estimation to the problem of cardinality estimation (see Section \ref{From Containment to Cardinality}).

\subsection{From Containment to Cardinality}
\label{From Containment to Cardinality}
Using a containment rate estimation models, we can obtain cardinality estimates using the Cnt2Crd transformation, as depicted in the rightmost diagram in Figure \ref{fig:teaser}.

\subsubsection{The Cnt2Crd Transformation}
\label{Cnt2Crd Transformation}
Given a containment rate estimation model\footnote{Here "model" may refer to an ML model or simply to a method.} $M$, we convert it to a cardinality estimation model using the Cnt2Crd transformation which returns a model $M'$ for estimating cardinalities. The obtained model $M'$ functions as follows. 
Given a "new" query, denoted as $Q_{new}$, as input to cardinality estimation, and assuming that there is an "old" query denoted as $Q_{old}$, whose FROM clause is the same as $Q_{new}$'s FROM clause, that has already been executed over the database, and therefore $|Q_{old}|$ is known, we:
\begin{itemize}
    \item Calculate $x\_rate = Q_{old} \subset_{\%} Q_{new}$ using $M$.
    \item Calculate $y\_rate = Q_{new} \subset_{\%} Q_{old}$ using $M$.
    \item Then, the cardinality estimate equals to: 
    $$ |Q_{new}| = \frac{x\_rate}{y\_rate} * |Q_{old}|$$
\end{itemize}
provided that $y\_rate = Q_{new} \subset_{\%} Q_{old}\  \neq 0$.

Given model $M$, we denote the obtained model $M'$, via the Cnt2Crd transformation, as Cnt2Crd($M$).

\begin{comment}

This is true, since:
$$x = Q_{old} \subset_{\%} Q_{new} = \frac{|Q_{new} \intersection Q_{old}|}{|Q_{old}|}$$
And,
$$y = Q_{new} \subset_{\%} Q_{old} = \frac{|Q_{new} \intersection Q_{old}|}{|Q_{new}|}$$
And therefore, 
$$ \frac{x}{y} = \frac{|Q_{new} \intersection Q_{old}|}{|Q_{old}|} * \frac{|Q_{new}|}{|Q_{new} \intersection Q_{old}|} = \frac{|Q_{new}|}{|Q_{old}|} $$

\end{comment}

\subsection{Queries Pool}
Our technique for estimating cardinality mainly relies on a queries pool that includes records of multiple queries.

The queries pool is envisioned to be an additional component of the DBMS, along with all the other customary components. It includes multiple queries with their actual cardinalities\footnote{Due to limited space, we do not detail the efficient hash-based data structures used to implement the queries pool.}, without the queries execution results. Therefore, holding such a pool in the DBMS as part of its meta information does not require significant storage space or other computing resources. Maintaining a queries pool in the DBMS is thus a reasonable expectation. The DBMS continuously executes queries, and therefore, we can easily configure the DBMS to store these queries along with their actual cardinalities in the queries pool.

In addition, we may generate in advance a queries pool using a queries generator that randomly creates multiple queries with many of the possible joins, and with different column predicates. We then execute these queries on the database to obtain and save their actual cardinalities in the queries pool.

Notice that, we can combine both approaches (actual computing and a generator) to create the queries pool. The advantage of the first approach is that in a real-world situation, queries that are posed in sequence by the same user, may be similar and therefore we can get more accurate cardinality estimates. The second approach helps in cases where the queries posed by users are diverse (e.g., different FROM clauses). Therefore, in such cases, we need to make sure, in advance, that the queries pool contains sufficiently many queries that cover all the possible cases.

Given a query $Q$ whose cardinality is to be estimated , it is possible that we fail to find any appropriate query, in the queries pool, to match with query $Q$. That happens when all the queries in the queries pool have a different FROM clause than that of query $Q$, or that they are not contained at all in query $Q$. In such cases we can always rely on the \textit{known} basic cardinality estimation models. In addition, we can make sure that the queries pool includes queries with the most frequent used FROM clauses, with empty column predicates. That is, queries of the following form.\\ \textit{SELECT * FROM -set of tables- WHERE TRUE}. In this case, for most of the queries posed in the database, there is at least one query that matches in the queries pool with the given query, and hence, we can estimate the cardinality without resorting to the basic cardinality estimation models.

\subsection{A Cardinality Estimation Technique}
\label{A Cardinality Estimation Technique}
Consider a new query $Q_{new}$, and assume that the DBMS includes a queries pool as previously described. To estimate the cardinality of $Q_{new}$ accurately, we use \textit{multiple} “old” queries instead of \textit{one} query, using the same Cnt2Crd transformation of Section \ref{Cnt2Crd Transformation}, as described in Figure \ref{Cardinality Estimation Technique}.

\begin{figure}[!h]
    \begin{mdframed}
    EstimateCardinality(Query $Q_{new}$, Queries Pool $QP$):\\
    \hspace*{0.4cm} $results$ = empty list\\ \\
    \hspace*{0.4cm} For every pair $(Q_{old},|Q_{old}|)$ in $QP$:\\
    \hspace*{0.8cm} if $Q_{old}$'s FROM clause $\neq$ $Q_{new}$'s FROM clause:\\
    \hspace*{1.2cm} continue\\ 
    \hspace*{0.8cm} Calculate $ x\_rate = Q_{old} \subset_{\%} Q_{new}$\\
    \hspace*{0.8cm} Calculate $ y\_rate = Q_{new} \subset_{\%} Q_{old}$\\
    \hspace*{0.8cm} if $y\_rate <= epsilon$:  /* y equals zero */ \\
    \hspace*{1.2cm} continue\\ 
    \hspace*{0.8cm} $results$.append($x\_rate/y\_rate * |Q_{old}|$)\\ \\
    \hspace*{0.4cm} return F($results$)
    \end{mdframed}
    \caption{Cardinality Estimation Technique.}
    \label{Cardinality Estimation Technique}
\end{figure}

Estimating cardinality considers all the \textit{matching} queries whose FROM clauses are identical to $Q_{new}$'s FROM clause. For each matching query, we estimate $Q_{new}$'s cardinality using the Cnt2Crd transformation and save the estimated result in the $results$ list. The final cardinality is obtained by applying the final function, $F$, that converts all the estimated results recorded in the $results$ list, into a single final estimation value. Note that the technique can be easily parallelized since each iteration in the For loop is independent, and thus can be calculated in parallel.

\subsubsection{Comparing Different Final Functions}
\label{Comparing Different Final Functions}

We examined various final functions ($F$), including:
\begin{itemize}
    \item Median, returning the median value of the $results$ list.
    \item Mean, returning the mean value of the $results$ list.
    \item Trimmed mean, returning the trimmed mean of the $results$ list without the 25\% outliers (trimmed mean removes a designated percentage of the largest and smallest values before calculating the mean).

\end{itemize}

Experimentally, the cardinality estimates using the various functions were very similar in terms of q-error. But the Median function yielded the best estimates (we do not detail these experiments due to limited space).

\section{Cardinality Evaluation}
\label{Cardinality Evaluation}
\label{Evalutaion Crd}
We evaluate our proposed technique for estimating cardinality, with different test sets, while using the CRN model as defined in Section \ref{Model} for estimating containment rates.

We compare our cardinality estimates with those of the PostgreSQL version 11 cardinality estimation component \cite{postgreSQL}, and the MSCN model \cite{LearnedCrd}. 

We train both the CRN model and the MSCN model with the \textit{same} training set as described in Section \ref{Comparing with MSCN}. Also, we create the test workloads using the same queries generator used for creating the training set of the CRN and the MSCN models (described in Section \ref{queries generator}), while skipping its last step. That is, we only run the first two steps of the generator. The third step creates query pairs which are irrelevant for the cardinality estimation task.

\subsection{Evaluation Workloads}
\label{Evaluation Workloads}
We evaluate our approach on the (challenging) IMDb dataset, using three different query workloads:
\begin{itemize}
    \item crd\_test1, a synthetic workload generated by the same queries generator that was used for creating the training data of the CRN model, as described in Section \ref{Defining the database and the development set} (using a different random seed) with 450 unique queries, with zero to two joins.
    \item crd\_test2, a synthetic workload generated by the same queries generator as the training data of the CRN model, as described in Section \ref{Defining the database and the development set} (using a different random seed) with 450 unique queries, with zero to \textit{five} joins. This dataset is designed to examine how the technique generalizes to additional joins.
    \item scale, another synthetic workload, with 500 unique queries, derived from the MSCN test set as introduced in \cite{LearnedCrd}. This dataset is designed to examine how the technique generalizes to queries that were \textit{not} created with the same trained queries' generator.
\end{itemize}

\begin{table}[h!]
\centering
\resizebox{\linewidth}{!}{
\begin{tabular}{lccccccc}
\hline
\hline
$\bold {number\ of\ joins}$ & $\bold {0}$ & $\bold {1}$ & $\bold {3}$ & $\bold {3}$ & $\bold {4}$ & $\bold {5}$ & $\bold {overall}$ \\
\hline
$\bold {crd\_test1}$ & $150$ & $150$ & $150$ & $0$ & $0$ & $0$ & $450$\\
$\bold {crd\_test2}$ & $75$ & $75$ & $75$ & $75$ & $75$ & $75$ & $450$\\
$\bold {scale}$ & $115$ & $115$ & $107$ & $88$ & $75$ & $0$ & $500$\\
\hline
\hline
\end{tabular}
}
\caption{Distribution of joins.}
\end{table}

\subsection{Queries Pool}
\label{Queries Pool}
Our technique relies on a queries pool, we thus created a synthetic queries pool, $QP$, generated by the same queries generator as the training data of the containment rate estimation model, as described in Section \ref{Defining the database and the development set} (using a different random seed) with 300 queries, equally distributed among all the possible FROM clauses over the database. In particular, $QP$, covers all the possible FROM clauses that are used in the tests workloads. Note that, there are no shared queries between $QP$ queries and the test workloads queries.

\begin{comment}
\begin{table}[h!]
\centering
\resizebox{\linewidth}{!}{
\begin{tabular}{lccccccc}
\hline
\hline
$\bold {number\ of\ joins}$ & $\bold {0}$ & $\bold {1}$ & $\bold {3}$ & $\bold {3}$ & $\bold {4}$ & $\bold {5}$ & $\bold {overall}$ \\
\hline
$\bold {QP}$ & $50$ & $50$ & $50$ & $50$ & $50$ & $50$ & $300$\\
\hline
\hline
\end{tabular}
}
\caption{Distribution of joins in $QP$.}
\end{table}
\end{comment}

Consider a query $Q$ whose cardinality needs to be estimated. On the one hand, the generated $QP$ contains "similar" queries to query $Q$, these can help the machine in predicting the cardinality. On the other hand, it also includes queries that are not similar at all to query $Q$, that may cause erroneous cardinality estimates. Therefore, the generated queries pool $QP$, faithfully represents a real-world situation.

\subsection{Experimental Environment}
\label{Experimenal Environment}
In all the following cardinality estimation experiments, for predicting the cardinality of a given query $Q$ in a workload $W$, we use the whole queries pool $QP$ as described in Section \ref{Queries Pool} with all its 300 queries. That is, the “old” queries used for predicting cardinalities, are the queries of $QP$. In addition, in all the experiments we use the Median function as the final $F$ function.

\subsection{The Quality of Estimates}
\begin{figure}[h!]
\begin{center}
  \includegraphics[width=\linewidth]{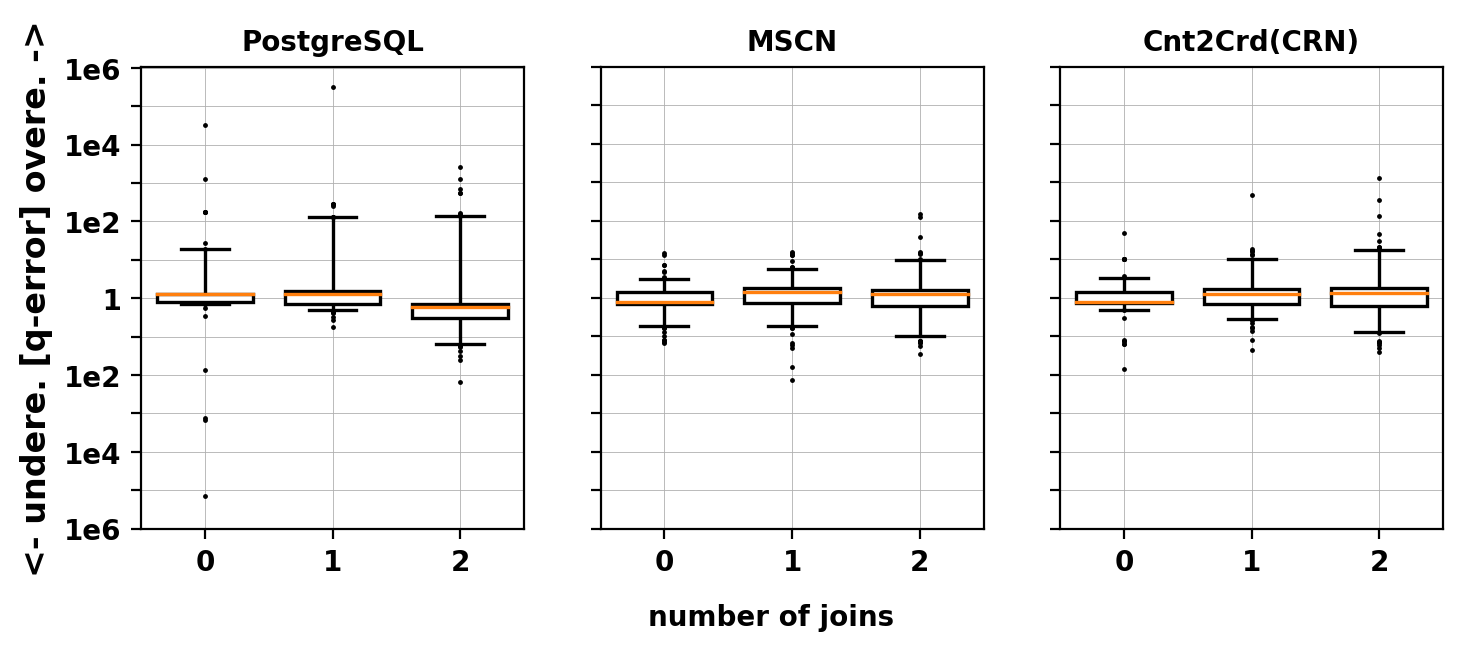}
  \caption{Estimation errors on the crd\_test1 workload.}
  \label{fig:crd_test1}
\end{center}
\end{figure}

Figure \ref{fig:crd_test1} depicts the q-error of the Cnt2Crd(CRN) model as compared to MSCN and PostgreSQL on  the crd\_test1 workload. While PostgreSQL's errors are more skewed towards the positive spectrum, MSCN is competitive with Cnt2Crd(CRN) in all the described values. As can be seen in Table \ref{table:crd_test1}, while MSCN provides the best results in the margins, the Cnt2Crd(CRN) model is more accurate in 75\% of the test. In addition, we show in the next section (Section \ref{Generalizing to additional joins-CRD}) that the Cnt2Crd(CRN) model is more robust when considering queries with more joins than in the training dataset.

\begin{table}[h!]
\centering
\resizebox{\linewidth}{!}{
\begin{tabular}{lccccccc}
\hline
\hline
$\bold {}$ & $\bold {50th}$ & $\bold {75th}$ & $\bold {90th}$ & $\bold {95th}$ & $\bold {99th}$ & $\bold {max}$ & $\bold {mean}$ \\
\hline
$\bold {PostgreSQL}$ & $1.74$ & $3.72$ & $22.46$ & $149$ & $1372$ & $499266$ & $1623$\\
$\bold {MSCN}$ & $2.11$ & $4.13$ & $\bold{7.79}$ & $\bold{12.24}$ & $\bold{51.04}$ & $\bold{184}$ & $\bold{4.66}$\\
$\bold{Cnt2Crd(CRN)}$ & $\bold{1.83}$ & $\bold{3.71}$ & $10.01$ & $18.16$ & $76.54$ & $1106$ & $9.63$\\
\hline
\hline
\end{tabular}
}
\caption{Estimation errors on the crd\_test1 workload.}
\label{table:crd_test1}
\end{table}

\newpage
\subsection{Generalizing to Additional Joins}
\label{Generalizing to additional joins-CRD}
We examine how our technique generalizes to queries with additional joins, without having seen such queries during training. To do so, we use the crd\_test2 workload which includes queries with zero to \textit{five} joins. Recall that we trained both the CRN model and the MSCN model only with query pairs that have between zero and two joins.

\begin{figure}[h!]
\begin{center}
  \includegraphics[width=\linewidth]{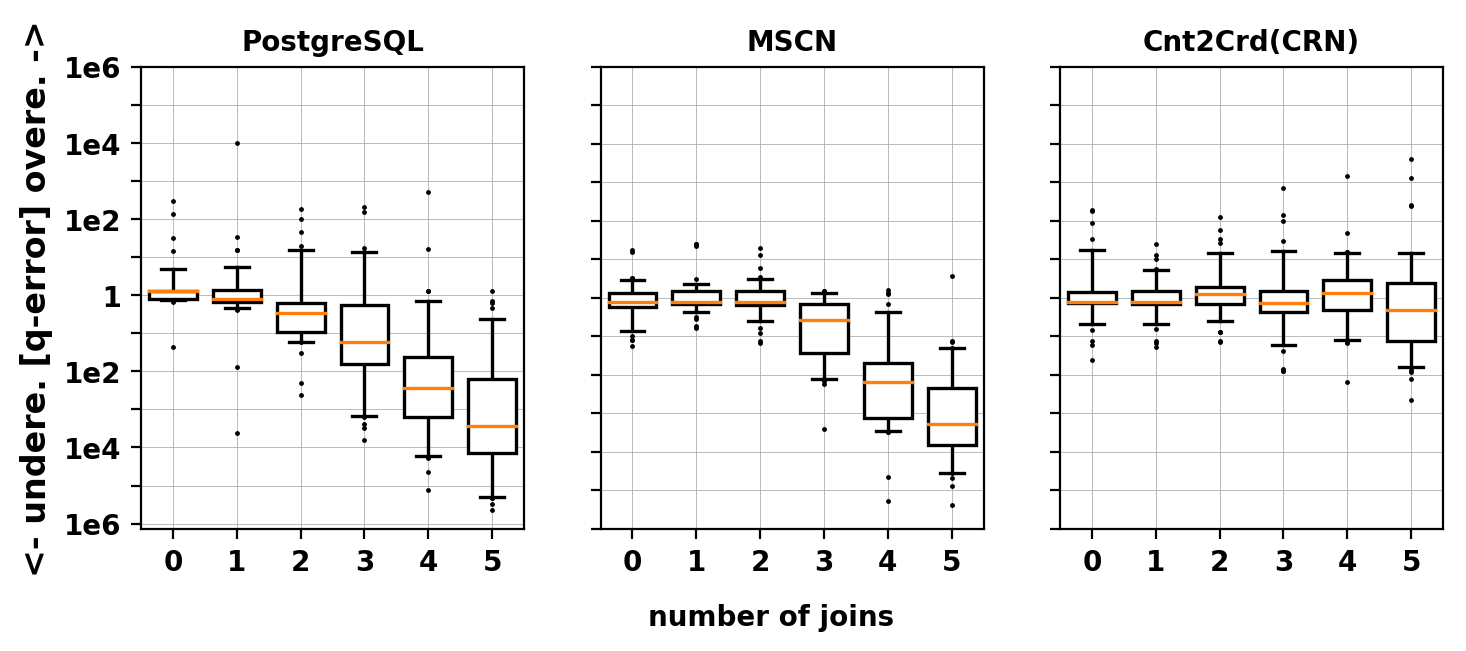}
  \caption{Estimation errors on the crd\_test2 workload.}
  \label{fig:crd_test2}
\end{center}
\end{figure}

From Tables \ref{table:crd_test2} and \ref{table:crd_test2joins3to5}, and Figure \ref{fig:crd_test2}, it is clear that Cnt2Crd(CRN) model is significantly more robust in generalizing to queries with additional joins. In terms of mean q-error, the Cnt2Crd(CRN) model reduces the mean by a factor x100 compared with MSCN and by a factor of x1000 compared with PostgreSQL.

\begin{table}[h!]
\centering
\resizebox{\linewidth}{!}{
\begin{tabular}{lccccccc}
\hline
\hline
$\bold {}$ & $\bold {50th}$ & $\bold {75th}$ & $\bold {90th}$ & $\bold {95th}$ & $\bold {99th}$ & $\bold {max}$ & $\bold {mean}$ \\
\hline
$\bold {PostgreSQL}$ & $9.22$ & $289$ & $5189$ & $21202$ & $576147$ & $4573136$ & $35169$\\
$\bold {MSCN}$ & $4.49$ & $119$ & $3018$ & $6880$ & $61479$ & $388328$ & $3402$\\
$\bold{Cnt2Crd(CRN)}$ & $\bold{2.66}$ & $\bold{6.50}$ & $\bold{18.72}$ & $\bold{72.74}$ & $\bold{528}$ & $\bold{6004}$ & $\bold{34.42}$\\
\hline
\hline
\end{tabular}
}
\caption{Estimation errors on the crd\_test2 workload.}
\label{table:crd_test2}
\end{table}

\begin{table}[h!]
\centering
\resizebox{\linewidth}{!}{
\begin{tabular}{lccccccc}
\hline
\hline
$\bold {}$ & $\bold {50th}$ & $\bold {75th}$ & $\bold {90th}$ & $\bold {95th}$ & $\bold {99th}$ & $\bold {max}$ & $\bold {mean}$ \\
\hline
$\bold {PostgreSQL}$ & $229$ & $3326$ & $22249$ & $166118$ & $2069214$ & $4573136$ & $70569$\\
$\bold {MSCN}$ & $121$ & $1810$ & $6900$ & $25884$ & $83809$ & $388328$ & $6801$\\
$\bold{Cnt2Crd(CRN)}$ & $\bold{4.28}$ & $\bold{10.84}$ & $\bold{43.71}$ & $\bold{93.11}$ & $\bold{1103}$ & $\bold{6004}$ & $\bold{61.26}$ \\
\hline
\hline
\end{tabular}
}
\caption{Estimation errors on the crd\_test2 workload considering only queries with three to five joins.}
\label{table:crd_test2joins3to5}
\end{table}

\begin{figure}[h!]
\begin{center}
  \includegraphics[width=\linewidth]{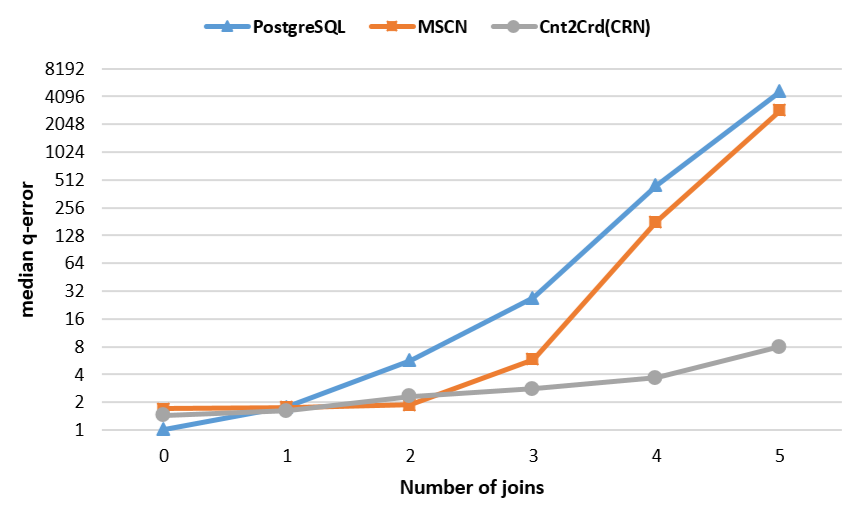}
  \caption{Q-error medians for each number of joins.}
  \label{fig:Q-error medians for each join}
\end{center}
\end{figure}

As depicted in Figure \ref{fig:crd_test2}, the Cnt2Crd(CRN) model generalizes more accurately to additional joins (note that the boxes are still on the same q-error interval). To highlight these improvements, we describe, in Table \ref{table:Q-error means for each join} and Figure \ref{fig:Q-error medians for each join}, the mean and median q-error for each possible number of joins separately (note the logarithmic y-axis scale in Figure \ref{fig:Q-error medians for each join}).
The known cardinality estimation models suffer from under-estimated results and errors that grow exponentially as the number of joins increases \cite{joinsHard}, as also happens in the cases we examined. The Cnt2Crd(CRN) model was better at handling additional joins (even though CRN was trained only with queries with up to two joins, as was MSCN).

\begin{table}[h!]
\centering
\resizebox{\linewidth}{!}{
\begin{tabular}{lcccccc}
\hline
\hline
$\bold {number\ of\ joins}$ & $\bold {0}$ & $\bold {1}$ & $\bold {2}$ & $\bold {3}$ & $\bold {4}$ & $\bold {5}$ \\
\hline
$\bold {PostgreSQL}$ & $10.41$ & $216$ & $25.38$ & $355$ & $4430$ & $210657$\\
$\bold {MSCN}$ & $\bold{3.44}$ & $3.56$ & $\bold{3.31}$ & $81.95$ & $5427$ & $14895$\\
$\bold{Cnt2Crd(CRN)}$ & $12.43$ & $\bold{3.54}$ & $6.77$ & $\bold{23.24}$ & $\bold{30.51}$ & $\bold{129}$\\
\hline
\hline
\end{tabular}
}
\caption{Q-error means for each number of joins.}
\label{table:Q-error means for each join}
\end{table}

\subsection{Generalizing to Different Kinds of Queries}

\begin{figure}[h!]
\begin{center}
  \includegraphics[width=\linewidth]{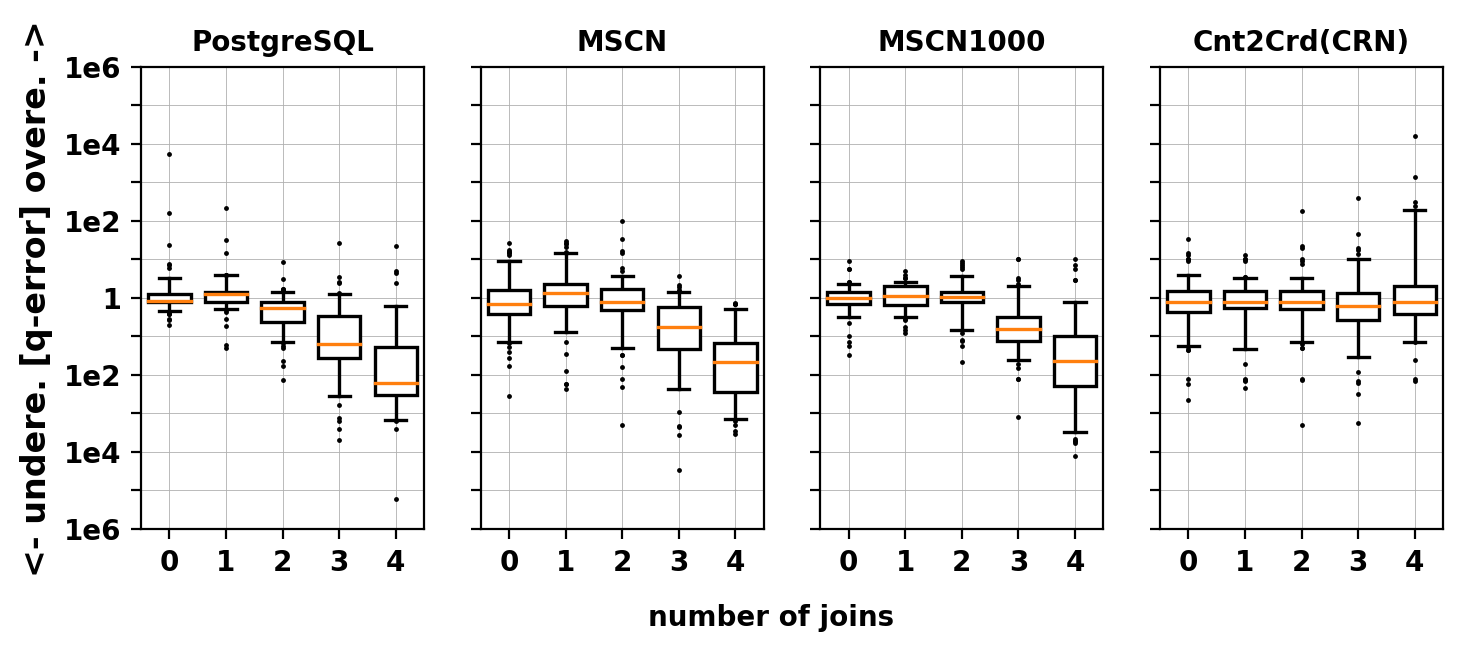}
  \caption{Estimation errors on the scale workload.}
  \label{fig:scaleB}
\end{center}
\end{figure}

In this experiment, we explore how the Cnt2Crd(CRN) model generalizes to a workload that was not generated by the same queries generator that was used for creating the CRN model training set. To do so, we examine the scale workload that was generated using another queries generator in \cite{LearnedCrd}. As shown in Table \ref{table:scale}, clearly Cnt2Crd(CRN) is more robust than MSCN and PostgreSQL in all the described values. Examining Figure \ref{fig:scaleB}, it is clear that the Cnt2Crd(CRN) model is significantly more robust with queries with 3 and 4 joins.
Recall that the $QP$ queries pool in this experiment was not changed, while the scale workload is derived from another queries generator. In summary, this experiment shows that Cnt2Crd(CRN) generalizes to workloads that were created with a different generator than the one used to create the training data.

\begin{table}[h!]
\centering
\resizebox{\linewidth}{!}{
\begin{tabular}{lccccccc}
\hline
\hline
$\bold {}$ & $\bold {50th}$ & $\bold {75th}$ & $\bold {90th}$ & $\bold {95th}$ & $\bold {99th}$ & $\bold {max}$ & $\bold {mean}$ \\
\hline
$\bold {PostgreSQL}$ & $2.62$ & $15.42$ & $183$ & $551$ & $2069$ & $233863$ & $586$\\
$\bold {MSCN}$ & $3.76$ & $16.84$ & $100$ & $448$ & $3467$ & $47847$ & $204$\\
$\bold{Cnt2Crd(CRN)}$ & $\bold{2.53}$ & $\bold{5.88}$ & $\bold{24.02}$ & $\bold{95.26}$ & $\bold{598}$ & $\bold{19632}$ & $\bold{69.85}$ \\
\hline
\hline
\end{tabular}
}
\caption{Estimation errors on the scale workload.}
\label{table:scale}
\end{table}

To further examine how Cnt2Crd(CRN) generalizes, we  conducted the following experiment. We compared the\\ Cnt2Crd(CRN) model with an improved version of the MSCN model that combines the deep learning approach and sampling techniques by using samples of 1000 materialized base tables, as described in \cite{LearnedCrd}. For simplicity we denote this model as MSCN1000. 

\begin{figure*}[h!]
\begin{center}
  \includegraphics[width=\linewidth]{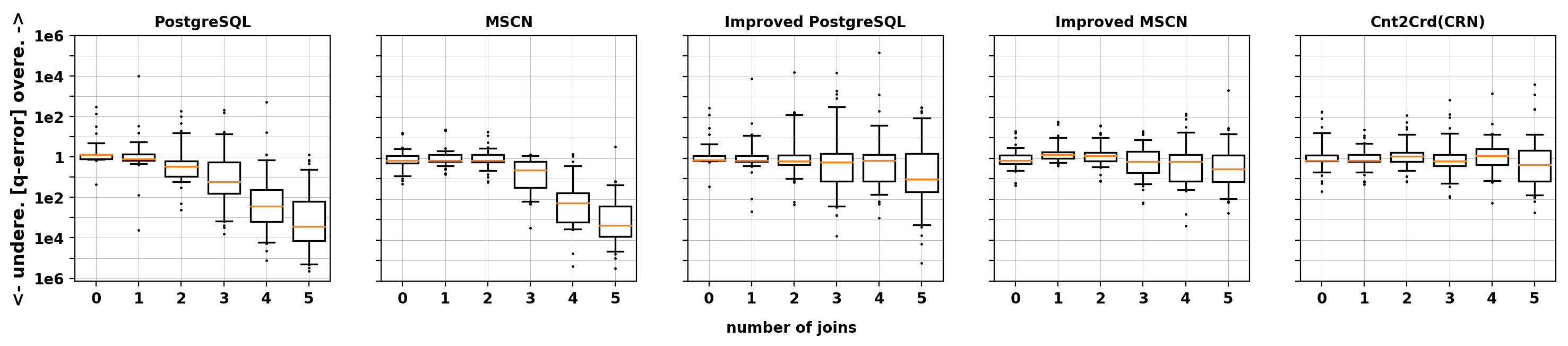}
    \caption{Estimation errors on the crd\_test2 workload, compared with all models.}
  \label{fig:all}
\end{center}
\end{figure*}

We make the test easier for MSCN1000 model by training the MSCN1000 model with a training set that was created with the \textit{same} queries generator that was used for generating the scale workload. As depicted in Figure \ref{fig:scaleB}, while the MSCN1000 model is more robust in queries with zero to two joins, still, the Cnt2Crd(CRN) model was found to be superior on queries with additional joins. Recall that the CRN model training set \textit{was not changed}, while the MSCN1000 model was trained with queries obtained from the \textit{same} queries generator that was used for creating the testing (i.e., scale) workload. In addition, note that MSCN1000 model uses sampling techniques whereas Cnt2Crd(CRN) does not. Thus, this experiment demonstrates the superiority of\\Cnt2Crd(CRN) in generalizing to additional joins.

\section{Improving Existing Cardinality Estimation Models}
\label{Improving Existing Cardinality Estimation Models}
In this section we describe how existing cardinality estimation models can be improved using the idea underlining our proposed technique. The proposed technique for improving existing cardinality estimation models relies on the same technique for predicting cardinalities using a containment rate estimation model, as described in Section \ref{A Cardinality Estimation Technique}.

In the previous section we used the CRN model in predicting containment rates. CRN can be replaced with \textit{any} other method for predicting containment rates. In particular, it can be replaced with any existing cardinality estimation model after "converting" it to estimating containment rates using the Crd2Cnt transformation, as described in Section \ref{From Cardinality to Containment}.

At first glance, our proposed technique seems to be more complicated for solving the problem of estimating cardinalities. However, we show that by applying it to known existing models, we improve their estimates, without changing the models themselves. These results indicate that the traditional approach, which directly addressed this problem, straightforwardly, using models to predict cardinalities, can be improved upon.

In the remainder of this section, we described the proposed approach, and show how existing cardinality estimation methods are significantly improved upon, by using this technique.

\newpage
\subsection{Approach Demonstration}
Given an existing cardinality estimation model $M$, we first convert $M$ to a model $M'$ for estimating containment rates, using the Crd2Cnt transformation, as described in Section \ref{From Cardinality to Containment}. Afterwards, given the obtained containment rate estimation model $M'$, we convert it to a model $M''$ for estimating cardinalities, using the Cnt2Crd transformation, as described in Section \ref{A Cardinality Estimation Technique}, which uses a queries pool. 

To summarize, our technique converts an existing cardinality estimation model $M$ to an intermediate model $M'$ for estimating containment rates, and then, using $M'$ we create a model $M''$ for estimating cardinalities with the help of the queries pool, as depicted in Figure \ref{fig:teaser} from left to right.

For simplicity, given cardinality estimation model $M$, we denote the model $M''$ described above, i.e., model\\ Cnt2Crd(Crd2Cnt($M$)), as \textit{Improved} $M$ model.

\subsection{Existing Models vs. Improved Models}
\label{Existing Models vs. Improved Models}

We examine how our proposed technique improves the PostgreSQL and the MSCN models, by using the crd\_test2 workload as defined in Section \ref{Evaluation Workloads}, as it includes the most number of joins. 
Table \ref{table:ImprovedPostgreSQL} depicts the estimates when using directly the PostgreSQL model, compared with the estimates when adopting our technique with PostgreSQL (i.e., the Improved PostgreSQL model). Similarly, Table \ref{table:ImprovedMSCN} depicts the estimates when using directly the MSCN model, compared with the Improved MSCN model. From both tables, it is clear that the proposed technique significantly improves the estimates (by factor x7 for PostgreSQL and x122 for MSCN in terms of mean q-error) without changing the models themselves (embedded within the Improved version).

These results highlight the power of our proposed approach that provides an effective and simple technique for improving existing cardinality estimation models. By adopting our approach and crating a queries pool in the database, cardinality estimates can be improved significantly.

\begin{table}[h!]
\centering
\resizebox{\linewidth}{!}{
\begin{tabular}{lccccccc}
\hline
\hline
$\bold {}$ & $\bold {50th}$ & $\bold {75th}$ & $\bold {90th}$ & $\bold {95th}$ & $\bold {99th}$ & $\bold {max}$ & $\bold {mean}$ \\
\hline
$\bold {PostgreSQL}$ & $9.22$ & $289$ & $5189$ & $21202$ & $576147$ & $4573136$ & $35169$\\
$\bold{Improved\ PostgreSQL}$ & $\bold{2.61}$ & $\bold{19.3}$ & $\bold{155}$ & $\bold{538}$ & $\bold{17697}$ & $\bold{1892732}$ & $\bold{5081}$ \\
\hline
\hline
\end{tabular}
}
\caption{Estimation errors on the crd\_test2 workload.}
\label{table:ImprovedPostgreSQL}
\end{table}

\begin{table}[h!]
\centering
\resizebox{\linewidth}{!}{
\begin{tabular}{lccccccc}
\hline
\hline
$\bold {}$ & $\bold {50th}$ & $\bold {75th}$ & $\bold {90th}$ & $\bold {95th}$ & $\bold {99th}$ & $\bold {max}$ & $\bold {mean}$ \\
\hline
$\bold {MSCN}$ & $4.49$ & $119$ & $3018$ & $6880$ & $61479$ & $388328$ & $3402$\\
$\bold{Improved\ MSCN}$ & $\bold{2.89}$ & $\bold{7.43}$ & $\bold{25.26}$ & $\bold{55.73}$ & $\bold{196}$ & $\bold{3184}$ & $\bold{27.78}$\\
\hline
\hline
\end{tabular}
}
\caption{Estimation errors on the crd\_test2 workload.}
\label{table:ImprovedMSCN}
\end{table}

\subsection{Improved Models vs. Cnt2Crd(CRN)}

Using the crd\_test2 workloade, we examine how our technique improves PostgreSQL and MSCN, compared with\\ Cnt2Crd(CRN). Adopting our technique improves the existing models as described in Section \ref{Existing Models vs. Improved Models}. Examining Table \ref{table:Improvedall}, it is clear that in 90\% of the tests, the best estimates are those obtained when directly using the CRN model to estimate the containment rates, instead of converting existing cardinality estimation models to obtain containment rates (Improved MSCN and Improved PostgreSQL).

\begin{table}[h!]
\centering
\resizebox{\linewidth}{!}{
\begin{tabular}{lccccccc}
\hline
\hline
$\bold {}$ & $\bold {50th}$ & $\bold {75th}$ & $\bold {90th}$ & $\bold {95th}$ & $\bold {99th}$ & $\bold {max}$ & $\bold {mean}$ \\
\hline
$\bold{Improved\ PostgreSQL}$ & $2.61$ & $19.3$ & $155$ & $538$ & $17697$ & $1892732$ & $5081$ \\
$\bold{Improved\ MSCN}$ & $2.89$ & $7.43$ & $25.26$ & $\bold{55.73}$ & $\bold{196}$ & $\bold{3184}$ & $\bold{27.78}$\\
$\bold{Cnt2Crd(CRN)}$ & $\bold{2.66}$ & $\bold{6.50}$ & $\bold{18.72}$ & $72.74$ & $528$ & $6004$ & $34.42$\\
\hline
\hline
\end{tabular}
}
\caption{Estimation errors on the crd\_test2 workload.}
\label{table:Improvedall}
\end{table}

\subsection{Cardinality Prediction Computation Time}
\label{Cardinality Prediction Time}
Using the proposed idea of using containment rates estimations to predict cardinalities, the cardinality prediction process is dominated by calculating the containment rates of the given input query with the relevant queries in the queries pool, and calculating the final function $F$ on these results to obtain the predicted cardinality, as described in Section \ref{A Cardinality Estimation Technique}. Therefore, the larger the queries pool is, the more accurate the predictions are, and the longer the prediction time is. Table \ref{table:poolSizes}, shows the medians and the means estimation errors on the crd\_test2 workload, along with the average prediction time for a single query, when using the Cnt2Crd(CRN) model for estimating cardinalities, with different sizes of $QP$ (equally distributed over all the possible FROM clauses in the database) while using the same final function $F$ (the Median function)).

\begin{table}[h!]
\centering
\resizebox{\linewidth}{!}{
\begin{tabular}{lcccccc}
\hline
\hline
$\bold{QP\ Size}$ & $\bold {50}$ & $\bold {100}$ & $\bold {150}$ & $\bold {200}$ & $\bold {250}$ & $\bold {300}$ \\
\hline
$\bold{Median}$ & $3.68$ & $2.55$ & $2.63$ & $2.55$ & $2.61$ & $2.66$ \\
$\bold{Mean}$ & $1894$ & $90$ & $41$ & $40$ & $35$ & $34$ \\
$\bold{Prediction\ Time}$ & 3.2ms & 7.1ms & 9.8ms & 11.3ms & 14.5ms & 16.1ms \\
\hline
\hline
\end{tabular}
}
\caption{Median and mean estimation errors on the crd\_test2 workload, and the average prediction time, considering different queries pool (QP) sizes.}
\label{table:poolSizes}
\end{table}

In table \ref{table:time}, we compare the average prediction time for estimating the cardinality of a single query using all the examined models (when using the whole $QP$ queries pool of size 300). While the default MSCN model is the fastest model, since it directly estimates the cardinalities without using a queries pool, the Cnt2Crd(CRN) model is the fastest among all the models that use a queries pool. That is, the Cnt2Crd(CRN) model is faster than the Improved MSCN model and the Improved PostgreSQL model. This is the case, since in the Improved MSCN model or the Improved PostgreSQL model, to obtain the containment rates, both models need to estimate cardinalities of two different queries as described in Section \ref{From Cardinality to Containment}, whereas the CRN model directly obtains a containment rate in one pass.

Although the prediction time of the models that use queries pools is higher than the most common cardinality estimation model (PostgreSQL), the prediction time is still in the order of a few milliseconds. In particular, it is similar to the average prediction time of models that use sampling techniques, such as the MSCN version with 1000 base tables samples. 

Recall that for the results in Table \ref{table:time}, we used a queries pool ($QP$) of size 300. We could have used a smaller pool, resulting in faster prediction time, and still obtaining better results using the models which employ a queries pool, as depicted in Table \ref{table:poolSizes}. Furthermore, all the the models that use queries pools may be easily parallelized as discussed in Section \ref{A Cardinality Estimation Technique}, and thus, reducing the prediction time (in our tests we ran these models serially).

\begin{table}[h!]
\centering
\resizebox{7cm}{!}{
\begin{tabular}{lc}
\hline
\hline
$\bold {Model}$ & $\bold {Prediction\ Time}$ \\
\hline
$\bold {PostgreSQL}$ & 2.1ms \\
$\bold {MSCN}$ & 1.1ms \\
$\bold {MSCN\ with\ 1000\ samples}$ & 33ms \\
$\bold {Improved\ PostgreSQL}$ & 70ms \\
$\bold{Improved\ MSCN}$ & 35ms \\
$\bold{Cnt2Crd(CRN)}$ & 16ms \\
\hline
\hline
\end{tabular}
}
\caption{Average prediction time of a single query.}
\label{table:time}
\end{table}

\section{Related Work}
\label{Related work}
Over the past five decades, conjunctive queries have been studied in the contexts of database theory and database systems. Conjunctive queries constitute a broad class of frequently used queries. Their expressive power is roughly equivalent to that of the Select-Join-Project queries of relational algebra. Therefore, several problems and algorithms have been researched in depth in this context. Chandra and Merlin \cite{ChandraMerlin} showed that determining containment of conjunctive queries is an NP-complete problem. Finding the minimal number of conditions that need to be added to a query in order to ensure containment in another query is also an NP-complete problem \cite{ulmanBook}. This also holds under additional settings involving inclusion and functional dependencies \cite{ulmanBook,foundationBook,dependencies}.

Although determining whether query $Q1$ is contained in query $Q2$ (analytically) in the case of conjunctive queries is an intractable problem in its full generality, there are many tractable cases for this problem. For instance, in \cite{Newcnt2,Newcnt22} it was shown that query containment in the case of conjunctive queries could be solved in linear time, if every database (edb) predicate occurs at most twice in the body of $Q1$. In \cite{Newcnt3} it was proved that for every $k \geq 1$, conjunctive query containment could be solved in polynomial time, if $Q2$ has querywidth smaller than $k+1$. In addition to the mentioned cases, there are many other tractable cases \cite{cnt1,cnt2,cnt3,cnt4}. Such cases are obtained by imposing syntactic or structural restrictions on the input queries $Q1$ and $Q2$. 

Whereas this problem was well researched in the past, to our knowledge, the problem of determining the containment rate on a \textit{specific} database has not been investigated. In this paper, we address this problem using ML techniques.

Lately, we have witnessed extensive adoption of machine learning, and deep neural networks in particular, in many different areas and systems, and in particular in databases. Recent research investigates machine learning for classical database problems such as join ordering \cite{MLjoinOrder}, index structures \cite{MLindex}, query optimization \cite{MLoptimiztion1,MLoptimiztion2}, concurrency control \cite{concurrency}, and recently in cardinality estimation \cite{LearnedCrd,LearnedCrd2}. In this paper, we propose a deep learning-based approach for predicting containment rates on a \textit{specific} database and show how containment rates can be used to predict cardinalities more accurately.

There were many attempts to tackle the problem of cardinality estimation; for example, Random Sampling techniques \cite{RS1,RS2}, Index based Sampling \cite{IBJS}, and recently deep learning \cite{LearnedCrd,LearnedCrd2}. However, all these attempts have addressed, conceptually, the problem directly in the same way, as a black box, where the input is a query, and the output is the cardinality estimate. We address this problem differently by using information about queries that have already been executed in the database, together with their actual result cardinalities, and the predicted containment rates between them and the new examined query. Using this new approach, we improve cardinality estimates significantly.

\section{Conclusions and Future Work}
\label{Conclusion}
We introduced a new problem, that of estimating containment rates between queries over a \textit{specific} database, and introduced the CRN model, a new deep learning model for solving it. We trained CRN with generated queries, uniformly distributed within a constrained space, and showed that CRN usually obtains the best results in estimating containment rates as compared with other examined models.

We introduced a novel approach for cardinality estimation, based on the CRN-based containment rate estimation model, and with the help of a queries pool. We showed the superiority of our new approach in estimating cardinalities more accurately than state-of-the-art approaches. Further, we showed that it addresses the weak spot of existing cardinality estimation models, which is handling multiple joins. 

In addition, we proposed a technique for improving \textit{any} existing cardinality estimation model without the need to change the model itself, by embedding it within a three step method. Given that the estimates of state-of-the-art models are quite fragile, and that our new approach for estimating cardinalities is simple, has low overhead, and is quite effective, we believe that it is highly promising and practical for solving the cardinality estimation problem. 

To make our containment based approach suitable for more general queries, the CRN model for estimating containment rates can be extended to support other types of queries, such as the union queries, and queries that include complex predicates. In addition, the CRN model can be configured to support databases that are updated from time to time. Next, we discuss some of these extensions, and sketch possible future research directions.

\textit{Strings.} 
 A simple addition to our current implementation may support equality predicates on strings. To do so, we could hash all the possible string literals in the database into the integer domain (similarly to MSCN). This way, an equality predicate on strings can be converted to an equality predicate on integers, which the CRN model can handle.

\textit{Complex predicates.} 
Complex predicates, such as LIKE, are not supported since they are not represented in the CRN model. To support such predicates we need to change the model architecture to handle such predicates. Note that predicates such as BETWEEN and IN, may be converted to ordinary predicates.

\textit{SELECT clause.} 
In this work we addressed only queries with a SELECT * clause. We can handle queries with SELECT clauses that include specific columns. Given such a query $Q$, $Q$'s cardinality is equivalent to the cardinality of the query with a SELECT * clause instead, as long as the DISTINCT keyword is not used.

\textit{EXCEPT Operator.} 
Given a query $Q$ of the form\\ $Q1$ EXCEPT $Q2$, the CRN model can handle $Q$. In terms of containment rates:
$$(Q1\ EXCEPT\ Q2) \subseteq_{\%} Q3 = Q1 \subseteq_{\%} Q3 -  (Q1 \intersection Q2) \subseteq_{\%} Q3$$
Using the same idea, we can handle the opposite containment direction case, and the case where there are more than two queries, recursively. The case when considering cardinalities is similar.
$$|Q1\ EXCEPT\ Q2| = |Q1| - |Q1 \intersection Q2|$$

\textit{Union Queries.} 
Given a query $Q$ of the from $Q1$ UNION $Q2$, the CRN model can handle $Q$ as follows:
$$(Q1\ UNION\ Q2) \subseteq_{\%} Q3$$
$$= Q1 \subseteq_{\%} Q3 +  Q2 \subseteq_{\%} Q3 - (Q1 \intersection Q2) \subseteq{\%} Q3$$
Using the same idea, we can handle the opposite containment direction case, and the case where there are more than two queries, recursively. The case when considering cardinalities is similar.
$$|Q1\ UNION\ Q2| = |Q1| +  |Q2|$$

\textit{The OR operator.}
Given queries that include the operator OR in their WHERE clause, the CRN model does not handle such queries straightforwardly. But, we can handle such queries using a recursive algorithm that converts the queries into multiple conjunctive queries by converting the WHERE clause to DNF, and considering every conjunctive clause as a separate query.
$$(Q1\ OR\ Q2) \subseteq_{\%} Q3 = (Q1\ UNION\ Q2) \subseteq_{\%} Q3$$
Using the same idea, we can handle the opposite containment direction case, and the case where there are more than two queries, recursively. The case when considering cardinalities is similar.
$$|Q1\ OR\ Q2| = |Q1\ UNION\ Q2| - |Q1 \intersection Q2|$$

\textit{Database updates.}
Thus far, we assumed that the database is static (read-only database). However, in many real world databases, updates occur frequently. In addition, the database schema itself may be changed. To handle updates we can use one of the following approaches: 

(1) We can always completely re-train the CRN model with a new updated training set. This comes with a considerable compute cost for re-executing queries pairs to obtain up-to-date containment rates and the cost for re-training the model itself. In this approach, we can easily handle changes in the database schema, since we can change the model encodings prior to re-training it.

(2) We can incrementally train the model starting from its current state, by applying new updated training samples, instead of re-training the model from scratch. While this approach is more practical, a key challenge here is to accommodate changes in the database schema. To handle this issue, we could hold, in advance, additional place holders in our model to be used for future added columns or tables. In addition, the values ranges of each column may change when updating the database, and thus, the normalized values may be modified as well. Ways to handle this problem are the subject of current research.

\begin{comment}
\begin{appendix}
\end{appendix}
\end{comment}

\balance
\bibliographystyle{abbrv}
\bibliography{main}

\end{document}